\documentclass[12pt,preprint]{aastex}

\usepackage{graphicx}
\usepackage{epsfig}
\usepackage{natbib}

\slugcomment{Submitted to the Astronomical Journal}
\shorttitle{XMM-Newton survey of the SDSS DR8} \shortauthors{Zhang et al.}

\received{2012}
\begin{document}

\title{Statistical Study of 2XMMi-DR3/SDSS-DR8 Cross-correlation Sample}

\author{Yan-Xia Zhang\altaffilmark{1}, Xin-Lin Zhou\altaffilmark{1}, Yong-Heng Zhao\altaffilmark{1} and Xue-Bing Wu\altaffilmark{2}}

\altaffiltext{1}{Key Laboratory of Optical Astronomy, National
Astronomical Observatories, Chinese Academy of Sciences, 20A Datun
Road, Chaoyang District, 100012, Beijing, P.R.China}
\altaffiltext{2}{Department of Astronomy, Peking University 100871,
Beijing, P.R.China} \email{zyx@bao.ac.cn}

\begin{abstract}
Cross-correlating the XMM-Newton 2XMMi-DR3 catalog with the Sloan
Digital Sky Survey (SDSS) Data Release 8, we obtain one of the
largest X-ray/optical catalogs and explore the distribution of
various classes of X-ray emitters in the multidimensional
photometric parameter space. Quasars and galaxies occupy different
zones while stars scatter in them. However, X-ray active stars have
a certain distributing rule according to spectral types. The earlier
the type of stars, the stronger X-ray emitting. X-ray active stars
have a similar distribution to most of stars in the $g-r$ versus
$r-i$ diagram. Based on the identified samples with SDSS spectral
classification, a random forest algorithm for automatic
classification is performed. The result shows that the
classification accuracy of quasars and galaxies adds up to more than
93.0\% while that of X-ray emitting stars only amounts to 45.3\%. In
other words, it is easy to separate quasars and galaxies, but it is
difficult to discriminate X-ray active stars from quasars and
galaxies.  If we want to improve the accuracy of automatic
classification, it is necessary to increase the number of X-ray
emitting stars, since the majority of X-ray emitting sources are
quasars and galaxies. The results obtained here will be used for the
optical spectral survey performed by the Large sky Area Multi-Object
fiber Spectroscopic Telescope (LAMOST, also named the Guo Shou Jing
Telescope), which is a Chinese national scientific research facility
operated by the National Astronomical Observatories, Chinese Academy
of Sciences.

\end{abstract}


\section{INTRODUCTION}

With large space-based and ground-based observational instruments
operating, various large sky survey projects have obtained huge
amounts of data in different bands. Examples include infrared
surveys (Two Micron All Sky Survey, UKIDSS, WISE), optical surveys
(USNO, SDSS), X-ray surveys (ROSAT, Chandra, XMM-Newton), radio
surveys (FIRST, NVSS), and so on. All of these provide chances to
study the multiwavelength properties of celestial objects. Combining
measurements from several instruments allows us to create spectral
energy distributions of celestial objects in a range of wavelengths
occupying a large part of the electromagnetic spectrum (e.g., Shang
et al. 2011). Multiwavelength studies may also lead to new
discoveries.  Lodieu et~al. (2012) cross-correlated the UKIRT
Infrared Deep Sky Survey (UKIDSS; Lawrence et al. 2007) and the
Sloan Digital Sky Survey Data Release 7 (SDSS DR7; Abazajian et al.
2009) and found new ultracool subdwarfs. Scholz (2010) used UKIDSS
data and their cross-correlation with SDSS data to obtain 11 new T
dwarf candidates. Multiwavelength data also contribute to the
improved accuracy of classification and photometric redshift
estimation. Laurino et~al. (2011) discussed a novel method called
Weak Gated Experts for extracting quasar candidates and determining
photometric redshifts. Y$\grave{e}$che et al. (2010) applied
artificial neural networks for quasar selection and photometric
redshift determination. Wu \& Jia (2010) performed quasar candidate
selection and photometric redshift estimation based on SDSS and
UKIDSS data. They proposed an empirical criterion in the $Y - K$
versus $g - z$ color-color diagram to separate stars and quasars
with redshift $z < 4$, and two other criteria for selecting
high-redshift quasars. Using the SDSS-UKIDSS color-redshift
relation, they estimated the photometric redshifts of 8498
SDSS-UKIDSS quasars, and found that a significant increase of the
photometric redshift accuracy was obtained compared to that based on
the SDSS color-redshift relation only.

It has been suggested that active galactic nuclei (AGNs, Seyfert
galaxies and quasars) contributed to the X-ray background (e.g.,
Comastri et~al. 1995). Thus X-ray sky surveys may be helpful in
detecting a significant population of AGNs. Deeper X-ray
extragalactic surveys performed by Chandra and XMM-Newton have
greatly contributed to our knowledge of the formation and evolution
of galaxies, clusters and groups of galaxies, and supermassive black
holes (e.g., Mushotzky et al. 2000; Hasinger et al. 2001). Such deep
surveys are pencil-beam, in that even the widest one only covers a
small part of the sky (see review by Brandt \& Hasinger 2005).
Wide-field survey in the COSMOS field covers an area of 2.13
deg$^2$, detected a total number of $\sim$2000 sources (Cappelluti
 et al. 2009). The all-sky hard-X-ray surveys are quite shallow, e.g.,
the 22 Month Swift/BAT all-sky survey detected only 461 sources
(Tueller et~al. 2010).

However, the XMM-Newton observatory provides unrivalled
 capabilities for serendipitous X-ray surveys with the largest effective
 area. The XMM-Newton serendipitous source catalog contains the
 largest number of sources ever obtained at X-ray energy (Watson et al. 2009).
 With the continual data
release of XMM-Newton and SDSS, we have the opportunity to obtain
one of the largest X-ray/optical samples covering a large sky area.
SDSS identifications of XMM-Newton sources provide accurate
photometric and spectroscopic properties of various X-ray emitters.
Pineau et~al. (2011) cross-identified 2XMMi catalog (Watson et~al.
2009) with SDSS DR7 (Abazajian et al. 2009), built an identified
sample, with this sample the way the various classes of X-ray
emitters gather in the multidimensional parameter space can be
analyzed and later used to design a source classification scheme.
Georgakakis \& Nandra (2011) selected 209 sources detected in the
2-8 keV spectral band with SDSS spectroscopic redshifts in the range
$0.03 < z < 0.2$ from the XMM-Newton survey of SDSS DR7 sample. Then
they explored the color-magnitude diagram of the sample and compared
it with that of X-ray-detected AGNs at $z\sim0.8$ in the AEGIS field
(Nandra et al. 2007). They found no evidence for evolution of the
X-ray AGN host colors from $z = 0.1$ to 0.8.

Here we present cross-identified 2XMMi-DR3 catalog with the SDSS DR8
(Aihara et~al. 2011) sample. Compared with previous work (e.g.,
Pineau et~al. 2011; Georgakakis \& Nandra 2011), we produce one of
the largest X-ray/optical samples over a large area. Note that only
a small part of the 2XMMi sample (3622/262,902$\approx 1.4\%$) is
covered by the SDSS spectral survey. We will use this sample as part
of the input catalog for the LAMOST (Large sky Area Multi-Object
fiber Spectroscopic Telescope) optical spectroscopic surveys (Zhao
et~al. 2012; Cui et~al. 2012). LAMOST (also named the Guo Shou Jing
Telescope) is a Chinese national scientific research facility
operated by the National Astronomical Observatories, Chinese Academy
of Sciences. The aperture of LAMOST is 4m, equipped  with 4000
automatic optical fibers. LAMOST has begun carrying out its optical
spectral survey of over 10 million stars and galaxies in the end of
2012 September.

This paper is organized as follows. Section 2 describes the
2XMMi-DR3 catalog, SDSS DR8 and the details of the
cross-identification of X-ray sources with SDSS DR8 optical objects,
provides the distribution of the main astrophysical classes of X-ray
emitters in the optical and X-ray parameter space and shows how
source classification could be done on this basis. Section 3
presents the principle of the random forest algorithm and applies
this algorithm to the automatic classification of X-ray emitters.
Section 4 summarizes this work and presents our conclusions.

\section{Catalogs}
\subsection{2XMMi-DR3 catalog}

The XMM-Newton satellite (Jansen et al. 2001) was launched by the
European Space Agency in late 1999. The further incremental version
of the Second XMM-Newton Serendipitous Source Catalog, 2XMMi-DR3
catalog (Watson et al. 2009) is the fifth publicly released
XMM-Newton X-ray source catalog produced by the XMM-Newton Survey
Science Center consortium on behalf of ESA. The 2XMMi-DR3 catalog,
released on 2010 April 28, has about 22\% new detections when
compared to the 2XMMi catalog and is the largest X-ray source
catalog ever produced, containing $\sim 3$ times as many discrete
sources as the ROSAT survey catalogs. 2XMMi-DR3 complements deeper
Chandra and XMM-Newton small-area surveys, probing a much larger sky
area. The catalogue provides important information for large samples
of various types of astrophysical objects including AGNs, clusters
of galaxies, interacting compact binaries and active stellar
coronae, with the power of X-ray selection. The large sky area
covered by the serendipitous survey also means that 2XMMi-DR3 is a
rich resource for exploring the variety of the X-ray source
populations and identifying rare source types. The catalog contains
353,191 X-ray source detections which relate to 262,902 unique X-ray
sources in the energy interval from 0.2 to 12 keV. The median flux
of the total photon energy band (0.2 - 12 keV) is $\sim 2.5 \times
10^{-14}$ erg s$^{-1}$ cm$^{-1}$; it is $\sim 5.6 \times 10^{-15}$
erg s$^{-1}$ cm$^{-1}$ in the soft energy band (0.2 - 2 keV), and it
is $\sim 1.4 \times 10^{-14}$ erg s$^{-1}$ cm$^{-1}$ in the hard
band (2 - 12 keV). About 20\% of the sources have total fluxes below
$1 \times 10^{-14}$ erg s$^{-1}$ cm$^{-1}$. The typical accuracy of
the source position is about 2 arcsec.

\subsection{SDSS DR8}

The SDSS (SDSS, York et~al. 2000) is an astronomical survey project,
which covers more than a quarter of the sky, to construct the first
comprehensive digital map of the universe in 3 three dimensions. A
large amount of spectroscopic and photometric data has been obtained
during the last several years by SDSS, which has opened a new
horizon for the study of galaxy properties such as galaxy evolution,
clusters, redshifts, large-scale distribution of morphological type,
and so on.

Following the Early Data Release and Data Releases 1-7 of SDSS-I/II,
SDSS-III will collect data from 2008 to 2014, using the 2.5m
telescope at Apache Point Observatory. The Eighth Data Release (DR8)
has been available since 2011 January. The release contains all of
the imaging data taken by the SDSS imaging camera (now totalling
over 14,000 deg$^{2}$), as well as new spectra taken by the SDSS
spectrograph during its last year of operation for the SEGUE-2
project. All of the imaging data have been reprocessed using
improved data processing pipelines. DR8 includes measurements for
nearly 500 million stars, galaxies, and quasars, and nearly two
million spectra. Building on the legacy of SDSS and SDSS-II, the
main scientific aim of the SDSS-III Collaboration is to map the
Milky Way, search for extrasolar planets, and solve the mystery of
dark energy. SDSS-III consists of four surveys (BOSS, SEGUE-2,
APOGEE, MARVELS), each focused on a different scientific theme.

\subsection{Cross-match}

Pineau et al. (2011) discussed the identification procedure between
the 2XMMi catalog and SDSS DR7 in detail. According to Pineau et al.
(2011), most SDSS counterparts with a 2XMMi identification
probability higher than $\sim$ 90\% are found less than 3 arcsec
from the X-ray position. Therefore in this work the cross-match
radius between 2XMMi-DR3 catalogue and SDSS DR8 is also set as 3
arcsec in order to keep higher identification possibility. Then the
nearest objects are taken as corresponding entries.
Cross-identifying the 2XMMi-DR3 catalogue with photometric archive
of SDSS DR8 and removing the default records, the number of entries
is 51,500. Then we discard SDSS entries with recorded magnitudes
fainter than 22.2 in any of the photometric bands considered. We
only study pointed sources with a positional error smaller than or
equal to 5 arcsec and further handle the data meeting errors on
$g-i<0.2$ to keep reliability of sample. Then the number of entries
in the 2XMM-SDSS photometric sample is 31,809.

Similarly, cross-matching the 2XMMi-DR3 catalog with the spectral
archive of SDSS DR8 and crossing out the default records, the number
of objects is 3622. Given the SDSS entries brighter than 22.2 mag,
and the errors on $g-r<0.2$ as well as limiting the positional error
smaller than or equal to 5 arcsec, the number of the 2XMM-SDSS
spectra sample becomes 3595. Detailed information about this sample
is indicated in Table 1. In Table~1, GALAXY, QSO and STAR are
adopted from SDSS DR8 spectra ``class", their subtypes are from SDSS
DR8 spectra `subclass'. BL, SB and SF are short for BROADLINE,
STARBURST and STARFORMING, respectively; G represents the objects
with ``class"=GALAXY and default ``subclass"; Q indicates the
objects with `class'=QSO and default `subclass'. The total number of
GALAXY, QSO, and STAR is 1358, 2120, and 117, separately.

\begin{table*}
\tiny
\centering
 \caption{The description of samples}
 \label{symbols}
 \begin{tabular}{l|ccccccccccccc}
  \hline
   \hline
  Total No.& &&&Subtype No.\\
 \hline
GALAXY&AGN (AGN BL) & BL& SB (SB BL) & SF (SF BL) & G   \\
1358&190 (49)& 96 &124 (3)& 257 (18)& 691\\
  \hline
QSO& AGN (AGN BL) & BL& SB (SB
BL) & SF (SF BL) & Q\\
2120&44 (40)&1704&171 (168)&15 (14)&186\\
  \hline
STAR&A&B&Carbon-lines&CV&F&G&K&L&M&OB&T&WD\\
117&6&1&2&15&13&5&13&2&53&3&2&2\\
 \hline
 \end{tabular}

 \end{table*}

In order to understand properties of X-ray sources, the various
distributions are presented in Figures~1-3 (see detailed
Figures~A1-A5 in the Appendix). Only from these figures are some
clustering characteristics obvious. As shown in Figures~1-3, the
pointed sources are easy to separate from extended sources. Quasars
are apparently discriminated from galaxies. However the subclasses
of galaxies and quasars are difficult to distinguish, especially for
subtypes of quasars. In the first left panel of Figure~1, the
objects seem to cluster in three parts. Comparing to the first right
panel of Figure~1, the three parts may correspond to quasars (left
cluster), galaxies (center cluster) and stars (lower right cluster).
In the top left panel of Figure~2, the objects seem to cluster in
two parts. Compared to the top right panel of Figure~2, the two
parts may correspond to quasars (upper cluster) and galaxies (lower
cluster). In the top left panel of Figure~3, the objects seem to
cluster in three parts. Comparing to the top right panel of
Figure~3, the three parts seemingly point to quasars (lower
cluster), galaxies (center cluster), and stars (upper cluster). Only
considering $g-i$ do galaxies likely have a sequence: SB, SF (AGN,
BL), G. The sequence gradually turns redder as $g-i$ increases.
Similarly, stars also have a sequence: CV, F, K, M, which becomes
redder as $g-i$ increases. These two sequences satisfy the physical
properties of galaxies and stars: the redder objects show a little
more X-ray emission. The X-ray emission turns fainter when $g-i$
becomes larger in the sequences. Given the top left, middle right
and bottom left panels in Figure~1, CV stars of all X-ray-emitting
stars are the strongest X-ray emitters and are easily confused with
quasars, and M stars are distinctly separated from quasars and
galaxies. From Figures~1-3, different subclasses of stars have a
distinct distribution, especially in the log($f_{\rm x}/f_{\rm r}$)
versus $g-i$ diagram and  the $g-i$ versus $hr2$ diagram. It is a
pity that the X-ray emitting star sample is too small to provide a
statistical conclusion.

Table~1 indicates that X-ray-emitting stars cover various stellar
spectral types including special stars (e.g., CVs, WDs). This means
that the X-ray emission in stars is not characteristic of a
particular class of stars. This is consistent with the viewpoint of
G\"udel \& Naz\'{e} (2009). They stated that stars located across
almost all regions of a Hertzsprung-Russell diagram have been
identified as X-ray sources, with only a few exceptions, most
notably A-type stars and the coolest M-type giants of spectral type
M. For cooler stars of F to M spectral classes, magnetic coronae,
overall similar to the solar corona, generate X-rays, enriched by
flares in which unstable magnetic fields reconnect and release
enormous amounts of energy in about minutes to hours. The presence
of coronae in these stars proves that the operation of an internal
dynamo generates the magnetic fields. Nevertheless, since the
fraction of X-ray emission stars is rather small in the whole
sample, we could not develop a common conclusion.

In order to study these X-ray active stars further, the distribution
of these stars in the $g-r$ versus $r-i$ diagram is shown in
Figure.~4. Comparing the stars with X-ray emitting and without X-ray
emitting, all stars in stripe 82 of SDSS DR8 are also plotted in
this figure. This is because stripe 82 is the region on the
celestial equator that SDSS has imaged over 10 times, giving coadded
optical data two times deeper than single epoch SDSS observations.
The magnitudes are dereddened according to the map of Schlegel
et~al. (1998). As shown in Figure.~4, the distribution of stars has
a turning point at nearly $g-r=1.3$. The different X-ray active
stars except M-type stars occupy the tilted horizontal branch while
M-type X-ray emitting stars lie in the tilt vertical branch. The
later stars have larger $r-i$. In particular, M stars have a rapid
increase with $g-r$ enhancement.

\section{Random forest for classification}
\subsection{Random Forest}
Random forest is an ensemble learning algorithm proposed by Brieman
(2001) that, given a set of class-labeled data, builds a set of
classification trees. Each tree is constructed from a bootstrap
sample of the training data. When constructing individual trees, an
arbitrary subset of attributes is chosen (hence the term ``random")
from which the best attribute for the split is selected. The
classification of new data points is based on the majority vote from
individually constructed tree classifiers in the forest. The
detailed steps are as follows.

1. Select a new bootstrap sample from the training set.

2. Grow an unpruned tree on this bootstrap.

3. At each internal node, randomly select $m_{\rm try}$ predictors
and determine the best split using only these predictors.

4. There is no need to perform cost complexity pruning. Save the
tree as is, alongside those built thus far. Output the overall
prediction according to majority vote from all individually trained
trees.

Random forests have some interesting properties (Breiman 2001). They
are more efficient, using some features in each node, instead of all
features. They also do not overfit as more trees are added.
Furthermore, they are relatively robust against outliers and noise
and they are easily parallelized. Random forests are often used when
we have very large training data sets and a very large number of
input variables (hundreds or even thousands of input variables).

Random forests have many successful applications in astronomy, for
example, supernova classification (Richards et~al. 2012), periodic
variable star classification (Dubath et~al. 2011; Richards et~al.
2011), multi-wavelength data classification (Gao et~al. 2009),
feature importance evaluation (Dubath et~al. 2011; Richards et~al.
2011), feature selection and feature weighting (Zhang et~al. 2010a,
2010b), and photometric redshift estimation (Carliles et~al. 2010).

\subsection{The classification result of
random forest}

We choose some parameters from SDSS-DR8 and 2XMMi-DR3 catalogs to
perform random forest. The chosen parameters are described in
Table~2. We have tried various input patterns, as shown in Table~3.
The best accuracy is 89.46\% with only optical information. With
only X-ray information, the best performance amounts to 75.10\%.
While using X-ray and optical information together, random forest
obtains a best performance of 92.71\% by a tenfold cross-validation
with the input pattern $hr2, hr3, hr4, r, r-i, g-r$, log$(f_{\rm
x}/f_{\rm r})$. The time taken to build the classified model costs
0.61 s. The result of this classification experiment is shown in
Table~4. The number of correctly classified instances is 3333, that
of incorrectly classified instances is 262, and each in the whole
sample achieves accuracies of 92.7\% and 7.3\%, respectively. The
accuracies of GALAXY, QSO and STAR is 93.4\%, 94.9\% and 45.3\%,
respectively. Apparently, GALAXY and QSO obtain satisfactory results
while STAR performs poorly. In other words, extragalactic sources
are easy to discriminate while X-ray-emitting stars are difficult to
separate from galaxies and quasars. Moreover, Table~4 shows that QSO
is easily classified as GALAXY, and GALAXY is also inclined to be
classified as QSO, but a few of them are misclassified as stars. The
main reason for the low accuracy for stars is the small star sample
compared to the number of galaxies and quasars. the imbalance of the
different samples necessarily influences the classification result.
The classification rule depends on the majority of the samples.
Naturally, the minor sample is as inclined toward misclassification
as the major sample. When we objectively double the star sample
(i.e., each star is input twice), the entire sample's accuracy
increases to 93.81\% and the accuracy of stars is 92.40\%. When star
sample is increased to 10 times (i.e. each star is input 10 times),
the entire sample's accuracy amounts to 95.32\% and the accuracy of
stars is 98.22\%. Obviously we hope that the classification
performance will improve with the increase of the star sample.
However, the number of X-ray-emitting stars is always smaller in
reality than galaxies and quasars. Consequently, an imbalance of
samples still exists, and thus the accuracy of stars continues to
grow with a larger star sample, though at a slower pace.

In addition, the performance of random forest compared with other
machine learning methods is explored. Briefly, the input pattern
($hr2, hr3, hr4, r, r-i, g-r$, log$(f_{\rm x}/f_{\rm r})$) is used
for different methods. The comparison result for different methods
is indicated in Table~5. Given the data and input pattern for our
case, random forest shows its superiority only in terms of accuracy.
The speed of constructing a model is the slowest for Support Vector
Machines (SVMs), and the fastest is for Radial Basis Function (RBF)
Network. SVMs show the poorest performance. Multi-Layer
Perceptron'performance is poorer than only random forest.
Considering both accuracy and speed in building a model, random
forest is the best choice. Usually, compared to SVMs and neural
networks, random forest has a low prediction accuracy and high
variance, but random forest has some advantages. For example, random
forest is more interpretable, feature importance can be estimated
during training for little additional computation, sample
proximities can be plotted, the visualization of output decision
trees can be supplied, random forest readily handles larger numbers
of predictors, the speed is faster to train, and random forest has
fewer model parameters and handles missing values, continuous and
categorical predictors, and problems where $k>>N$. Given our case,
random forest is superior to SVMs and neural networks. In other
words, random forest performs better than SVMs and neural networks
when the training sample is smaller.

\begin{table*}[ht]
\begin{center}
\caption{The chosen parameters, definition, catalogues and wavebands
}
\begin{tabular}{rlll}
\hline \hline
Parameter&Definition  &Catalogue& Waveband\\
\hline
$g$  & $g$ magnitude&SDSS& Optical band\\
$r$  & $r$ magnitude&SDSS& Optical band\\
$i$  & $i$ magnitude&SDSS& Optical band\\
$hr2$&Hardness ratio 2 &XMM&X-ray band\\
    &Definition: $hr2=(B-A)/(B+A)$, where&&\\
    &A=countrate in energy band 0.5-1keV&&\\
    &B=countrate in energy band 1-2keV&&\\
$hr3$&Hardness ratio 3 &XMM&X-ray band\\
    &Definition: $hr3=(C-B)/(C+B)$, where&&\\
    &B=countrate in energy band 1-2keV&&\\
    &C=countrate in energy band 2-4.5keV&&\\
$hr4$&hardness ratio 4 &XMM&X-ray band\\
    &Definition: $hr4=(D-C)/(D+C)$, where&&\\
    &C=countrate in energy band 2-4.5keV&&\\
    &D=countrate in energy band 4.5-12keV&&\\
SC\_EXTENT&Source extent&XMM&X-ray band\\
log$(f_{\rm x}/f_{\rm r})$&X-ray-to-optical flux ratio&SDSS,XMM&Optical and X-ray bands\\
\hline \hline
\end{tabular}
\bigskip
\end{center}
\end{table*}

\begin{table*}
\begin{center}
\caption{Accuracy with different input patterns}
\begin{tabular}{rlll }
\hline \hline
Input pattern& Accuracy \\
\hline
$r-i,g-r,r$                               &88.93\%\\
$g-i,r-i,g-r,g,r$                         &\textbf{89.46\%}\\
$hr2,hr3,hr4,log(f_{\rm x})$              &71.62\%\\
$hr2,hr3,hr4,log(f_{\rm x}),$SC\_EXTENT   &\textbf{75.10\%}\\
$hr2,g-i,r-i,g-r,log(f_{\rm x}/f_{\rm r})$&91.99\%\\
$hr2,hr3,hr4,g,r,g-i,r-i,g-r,log(f_{\rm x}/f_{\rm r})$&92.43\%\\
$hr2,hr3,hr4,g,r,g-i,r-i,g-r,log(f_{\rm x}/f_{\rm r}),$SC\_EXTENT& 92.63\%\\
$hr2,hr3,hr4,g,r,g-i,r-i,log(f_{\rm x}/f_{\rm r}),$SC\_EXTENT& 91.99\%\\
$hr2,hr3,hr4,g,r,g-i,r-i,log(f_{\rm x}/f_{\rm r})$& 92.04\%\\
$hr3,hr4,g,r,g-i,r-i,log(f_{\rm x}/f_{\rm r})$& 92.29\%\\
$hr2,hr3,hr4,r,r-i,g-r,log(f_{\rm x}/f_{\rm r}),$SC\_EXTENT& 92.65\%\\
$hr2,hr3,hr4,r,r-i,g-r,log(f_{\rm x}/f_{\rm r})$&\textbf{92.71\%}\\
\hline \hline
\end{tabular}
\bigskip
\end{center}
\end{table*}

\begin{table*}
\begin{center}
\caption{The classification result by random forest}
\begin{tabular}{rlll }
\hline \hline
classified$\downarrow$known$\to$& GALAXY &QSO&STAR\\
\hline
        GALAXY&1268 &  104  &33\\
        QSO   &83   &  2012 &31\\
        STAR  & 7   &  4    &53\\
\hline
       accuracy & 93.4\% & 94.9\%&45.3\% \\
\hline
\end{tabular}
\bigskip
\end{center}
\end{table*}

\begin{table*}
\begin{center}
\caption{The performance comparison for different methods}
\begin{tabular}{rlll }
\hline \hline
Method& Accuracy &Time taken to build model (seconds)\\
\hline
Radial Basis Function Network &90.88\% & 0.53 \\
Support Vector Machines     &86.56\%   & 29.64 \\
Multi-Layer Perceptron        & 91.82\%  &7.91  \\
Random Forest&92.71\%&0.61\\
 \hline \hline
\end{tabular}
\bigskip
\end{center}
\end{table*}

\section{Conclusions}

We cross-correlate the 2XMMi-DR3 catalog containing 262902 unique
X-ray sources with Data Release 8 of SDSS including images for
nearly 500 million stars, galaxies, and quasars, and spectra for
nearly two million sources. Since most SDSS entries with a 2XMMi
identification probability higher than $\sim$90\% are less than 3
arcsec from the X-ray position (Pineau et~al. 2011), the cross-match
error radius is set to 3 arcsec. The standard processing of data is
similar to that of Pineau et~al. (2011), so we can see whether or
not there is improvement when the data is increased. Due to more
detailed spectral information of stars provided by SDSS DR8 than by
SDSS DR7, X-ray properties of various spectral types of stars are
easy to study. Figures~1-3 show that extragalactic sources clusters,
galaxies and quasars have different distributions, and stars are
difficult to separate from extragalactic sources, but nevertheless,
different types of stars have some distribution sequence, especially
in Figures.~1 and 3. Of X-ray active stars, CV and WD stars are
stronger X-ray sources. The X-ray emission of other stars from
early-type to late-type becomes weaker. Figure.~4 also indicates
that X-ray-emitting stars have a similar distribution to most stars.
Quasars and CV stars are stronger X-ray emitters. M stars are easily
discriminated from other X-ray emitters. The random forest algorithm
is applied to the cross-matched sample. The classification result
further proves that quasars and galaxies are easy to distinguish
from each other and X-ray-emitting stars are apt to be misclassified
as galaxies or quasars. Quasars and galaxies are seldom classified
as stars. Due to the small number of stars in this work, we only
present a preliminary result. A more accurate and quantitative
conclusion will be obtained with more data from X-ray emitting stars
collected in future work. According to the distribution of
photometric data and spectral data as well as classifiers created by
random forest, it is convenient to select out quasar candidates and
galaxy candidates. This is important in order to build complete
samples of quasars or galaxies for statistical study. In addition,
the classification results from different approaches are compared.
When lacking a sufficiently trained sample, random forest shows its
superiority.

\acknowledgements{We are very grateful for the referee's
constructive suggestions and comments that helped to improve our
paper. This paper is funded by the National Natural Science
Foundation of China under grant Nos.10778724, 11178021, 11033001 and
No.11003022(XLZ), the Natural Science Foundation of the Education
Department of Hebei Province under grant No. ZD2010127 and by the
Young Researcher Grant of the National Astronomical Observatories,
Chinese Academy of Sciences. We acknowledge the SDSS database.
Funding for SDSS-III has been provided by the Alfred P. Sloan
Foundation, the Participating Institutions, the National Science
Foundation, and the U.S. Department of Energy Office of Science. The
SDSS-III Web site is http://www.sdss3.org/. SDSS-III is managed by
the Astrophysical Research Consortium for the Participating
Institutions of the SDSS-III Collaboration including the University
of Arizona, the Brazilian Participation Group, Brookhaven National
Laboratory, University of Cambridge, Carnegie Mellon University,
University of Florida, the French Participation Group, the German
Participation Group, Harvard University, the Instituto de
Astrofisica de Canarias, the Michigan State/Notre Dame/JINA
Participation Group, Johns Hopkins University, Lawrence Berkeley
National Laboratory, Max Planck Institute for Astrophysics, Max
Planck Institute for Extraterrestrial Physics, New Mexico State
University, New York University, Ohio State University, Pennsylvania
State University, University of Portsmouth, Princeton University,
the Spanish Participation Group, University of Tokyo, University of
Utah, Vanderbilt University, University of Virginia, University of
Washington, and Yale University.}

\clearpage

\begin{figure}
\includegraphics[bb=11 16 509 522,width=6cm,clip]{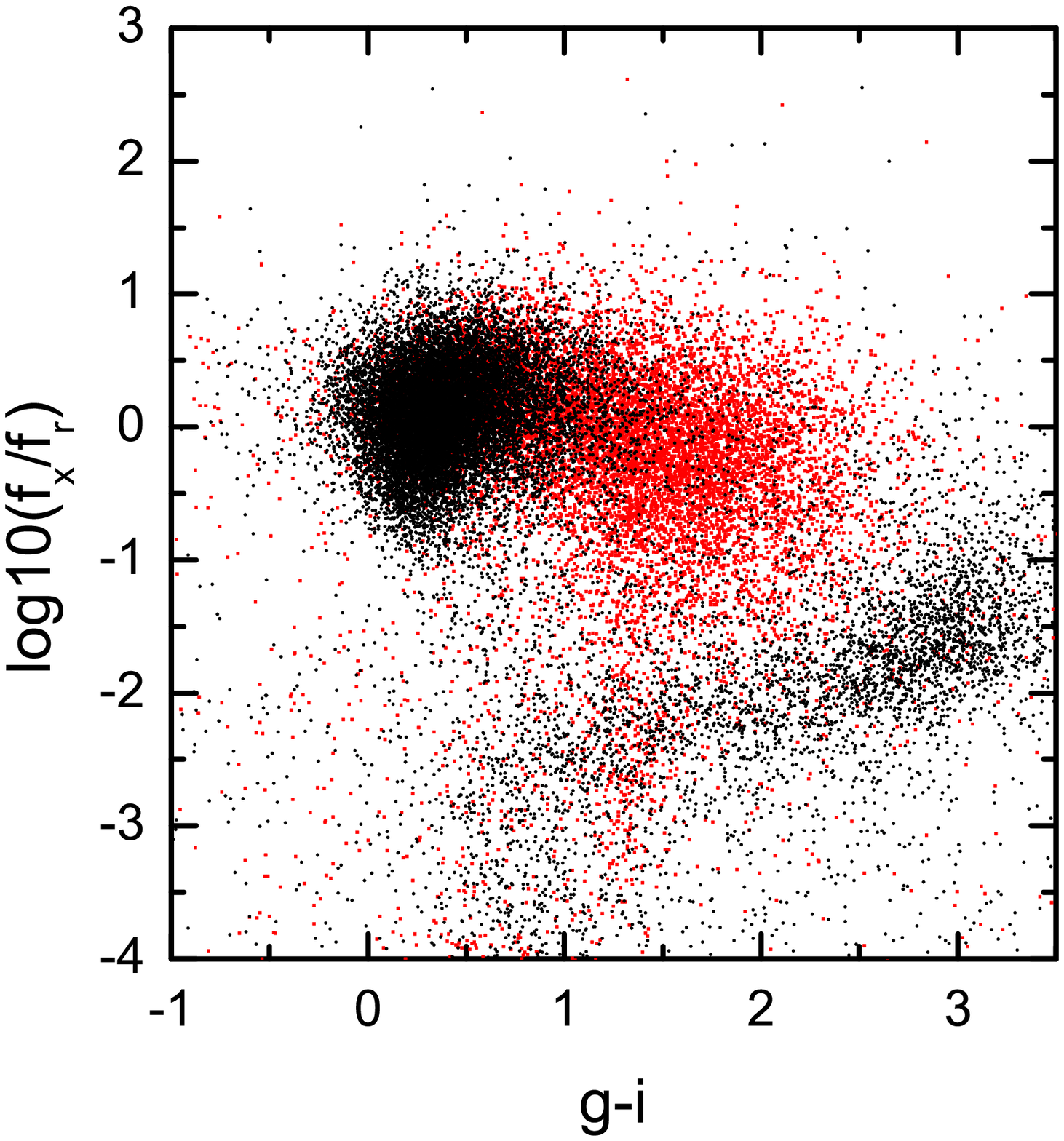}
\includegraphics[bb=11 16 509 522,width=6cm,clip]{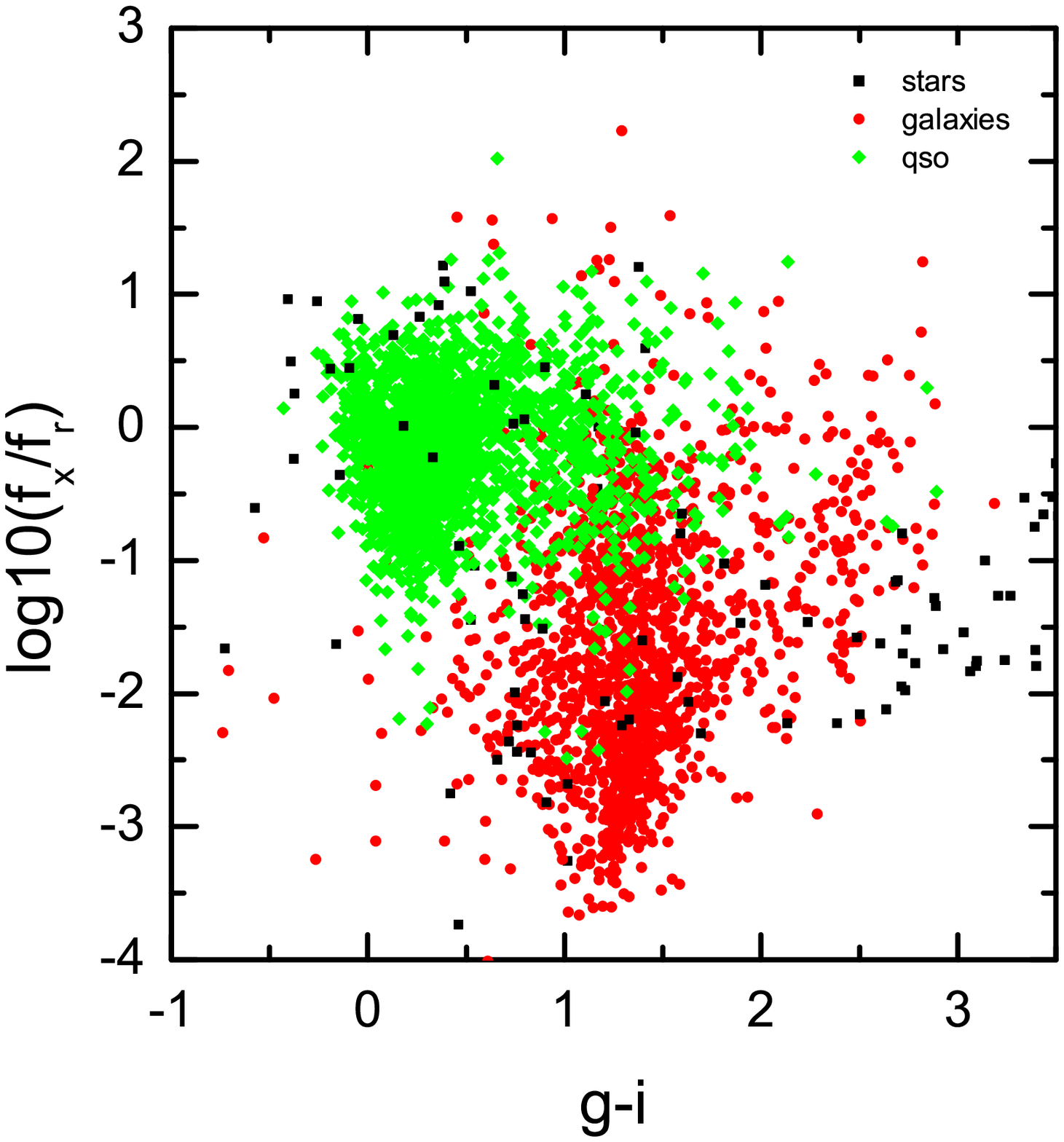}
\includegraphics[bb=11 16 509 522,width=6cm,clip]{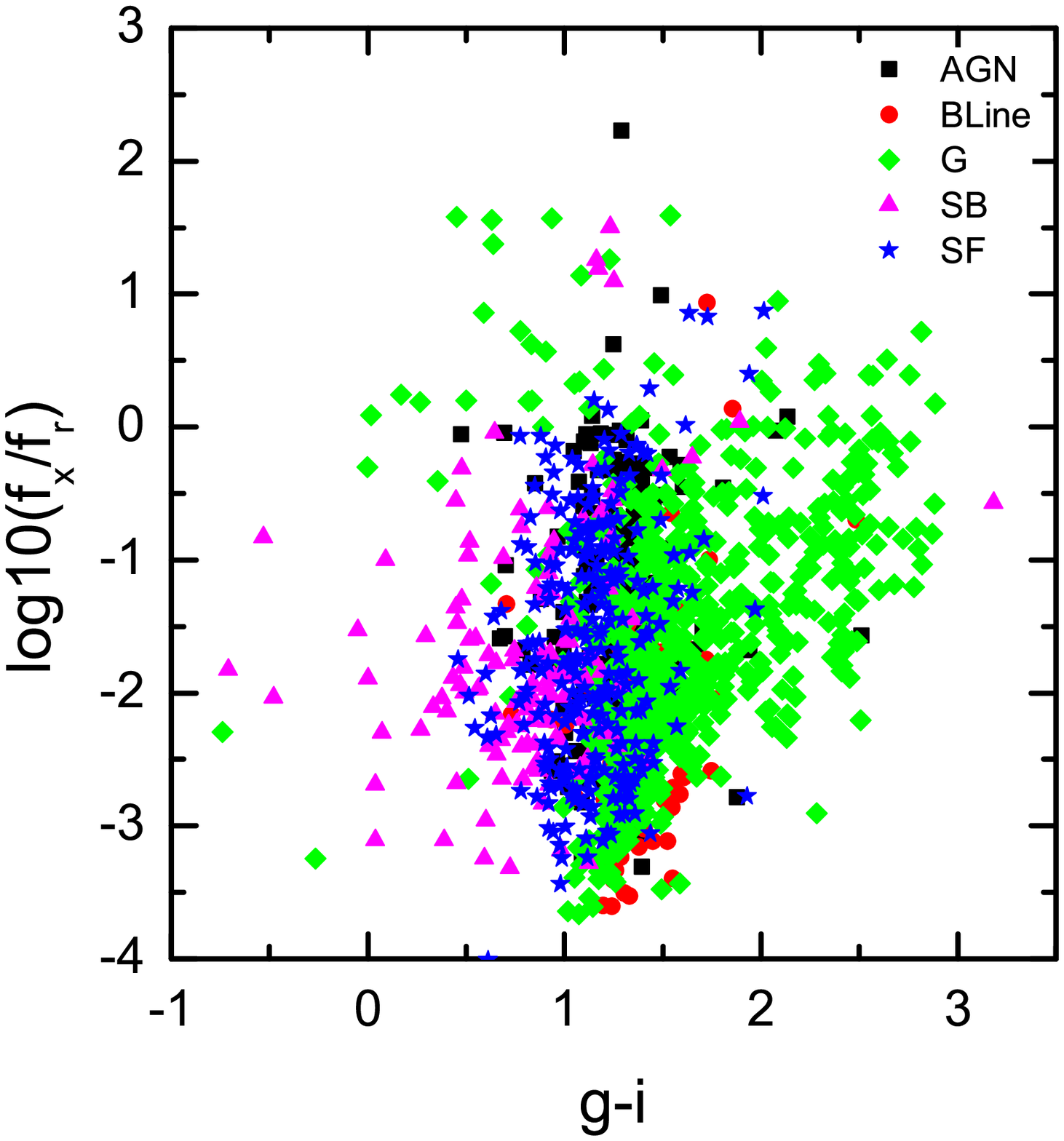}
\includegraphics[bb=11 16 509 522,width=6cm,clip]{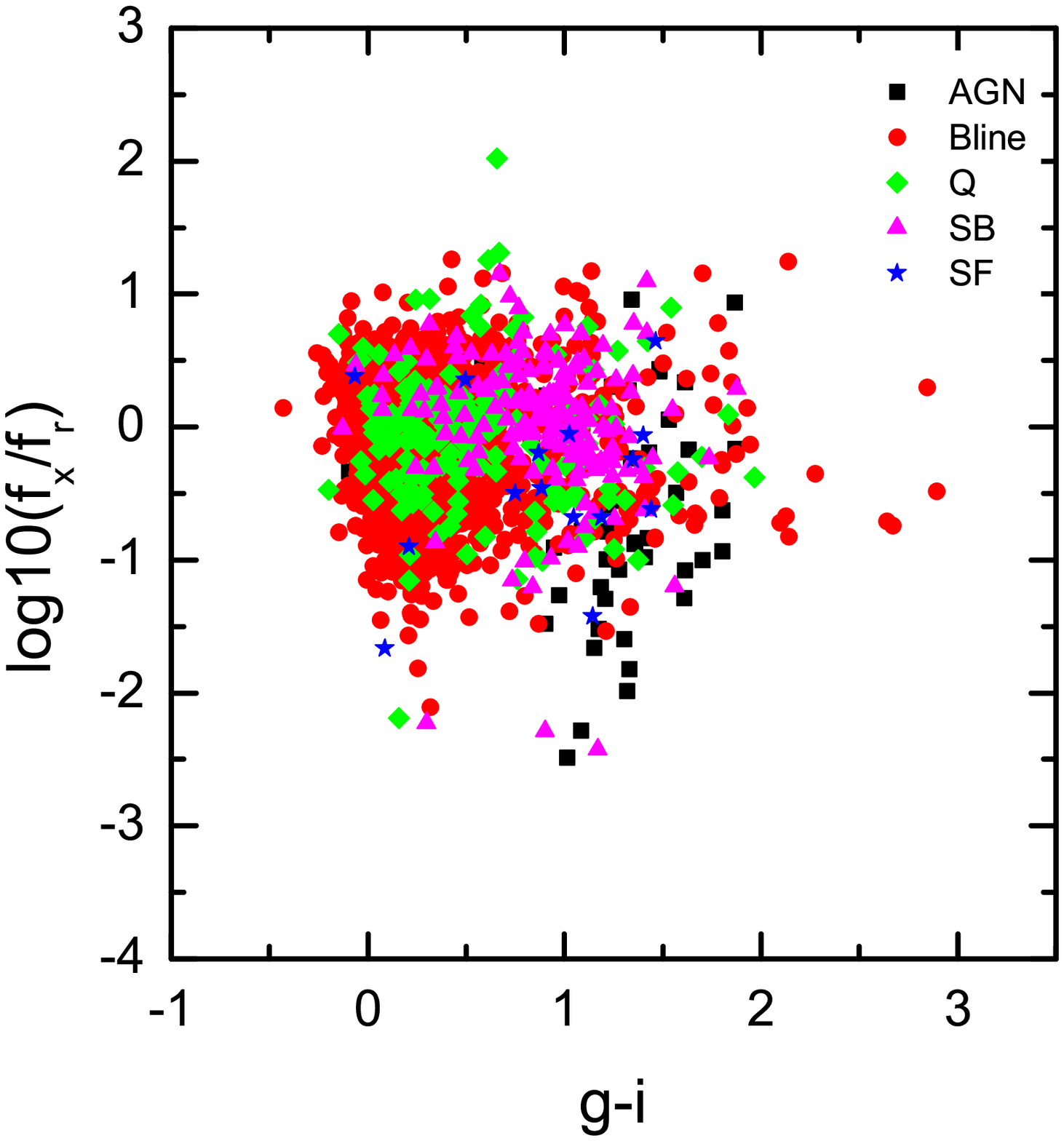}
\includegraphics[bb=11 16 509 522,width=6cm,clip]{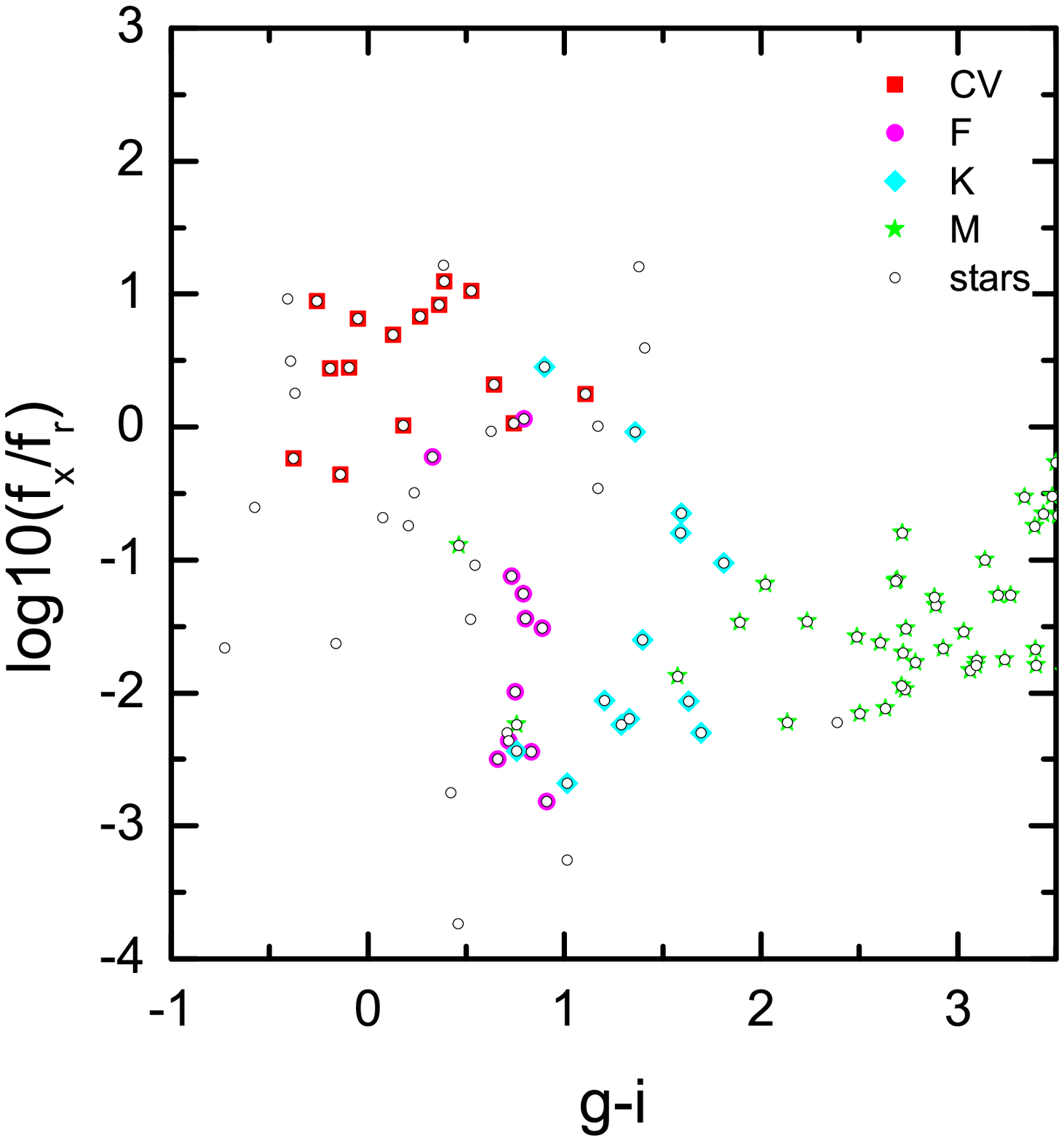}
  \caption{Distribution of sources in the optical band in the log($f_{\rm x}/f_{\rm r}$) versus $g-i$ diagram.
  Left No.1: the entire SDSS photometric sample. Pointed sources (black filled circles) and extended sources
  (red filled squares) point to the objects labeled as `STAR' and `GALAXY' in SDSS photometric archive, respectively.
  Right No.1: the objects with identified spectra. Stars are represented as black filled squares, galaxies as red filled
  circles, qso as green diamonds. Left No.2: the galaxy sample. Black filled squares: AGN, red filled circles: BL, green filled diamonds: G, magenta filled triangles: SB, cyan stars: SF. Right No.2: the quasar sample. Black filled squares: AGN, red filled circles: BL, green filled diamonds: Q, magenta filled triangles: SB, cyan stars: SF. Left No.3: the star sample. The opened circles: the whole star sample, red filled squares: CV, magenta filled circles: F stars, cyan filled diamonds: K stars, filled green stars: M stars. }
\end{figure}

\begin{figure}
\includegraphics[bb=11 16 509 522,width=6cm,clip]{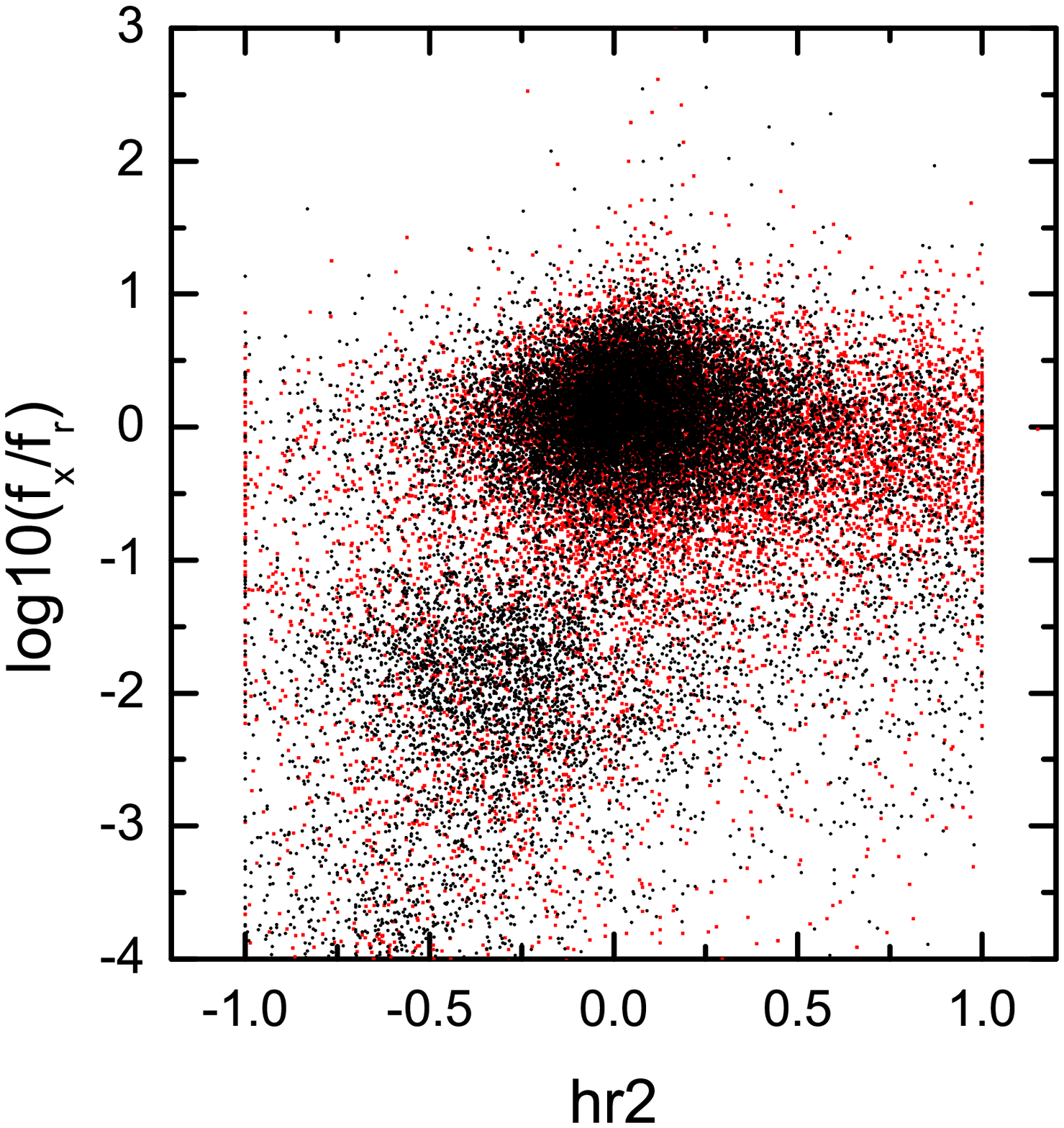}
\includegraphics[bb=11 16 509 522,width=6cm,clip]{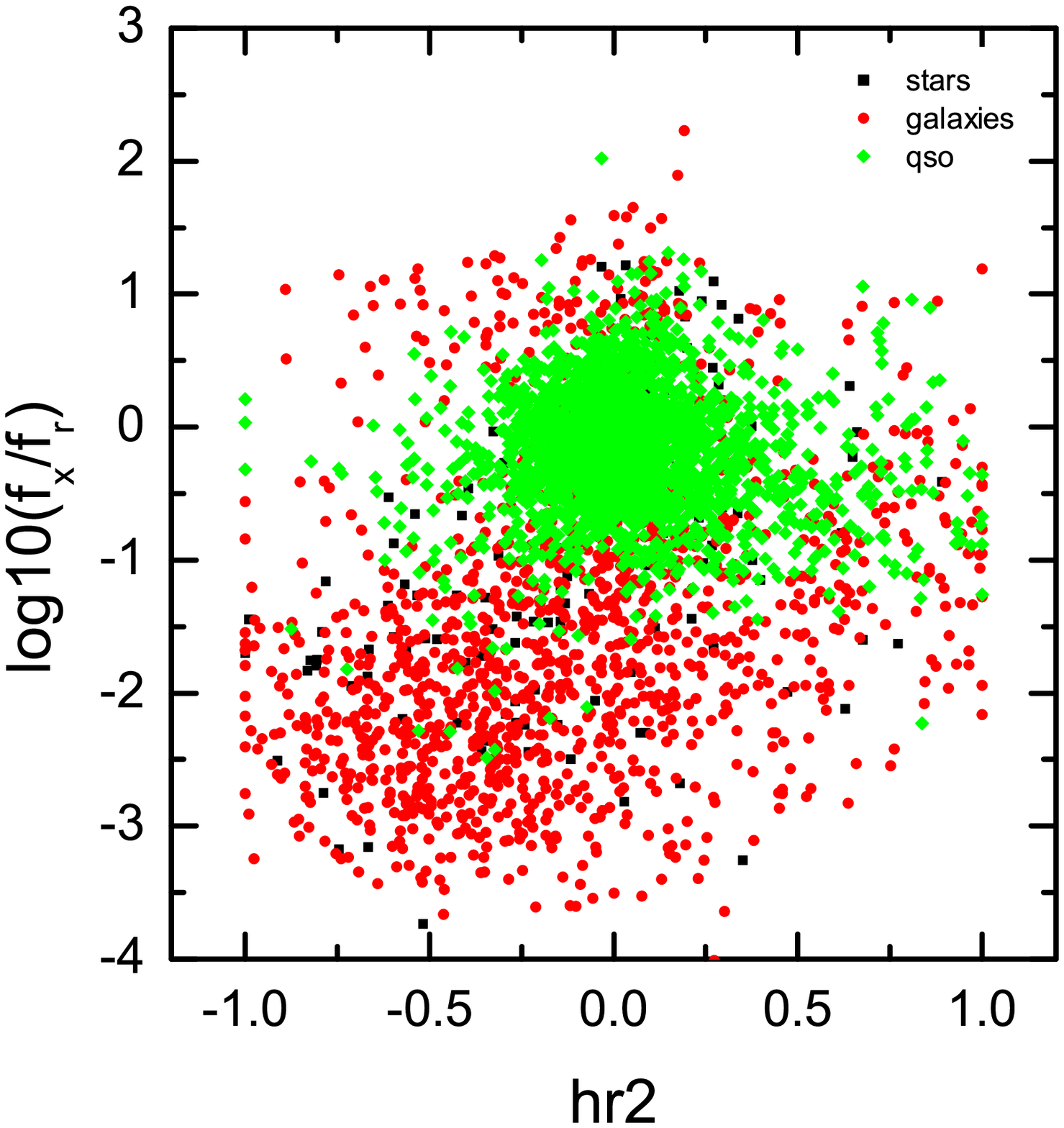}
\includegraphics[bb=11 16 509 522,width=6cm,clip]{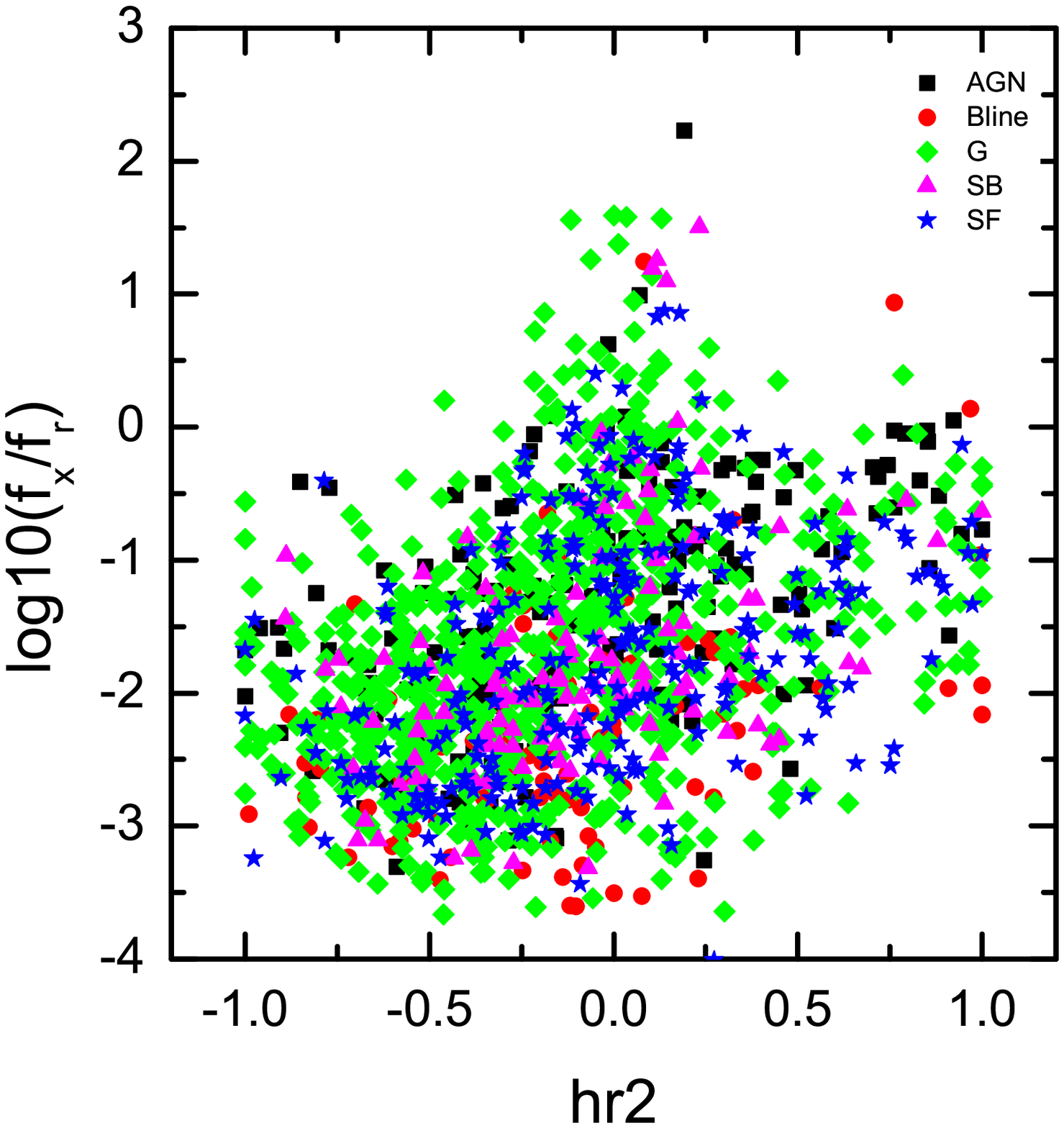}
\includegraphics[bb=11 16 509 522,width=6cm,clip]{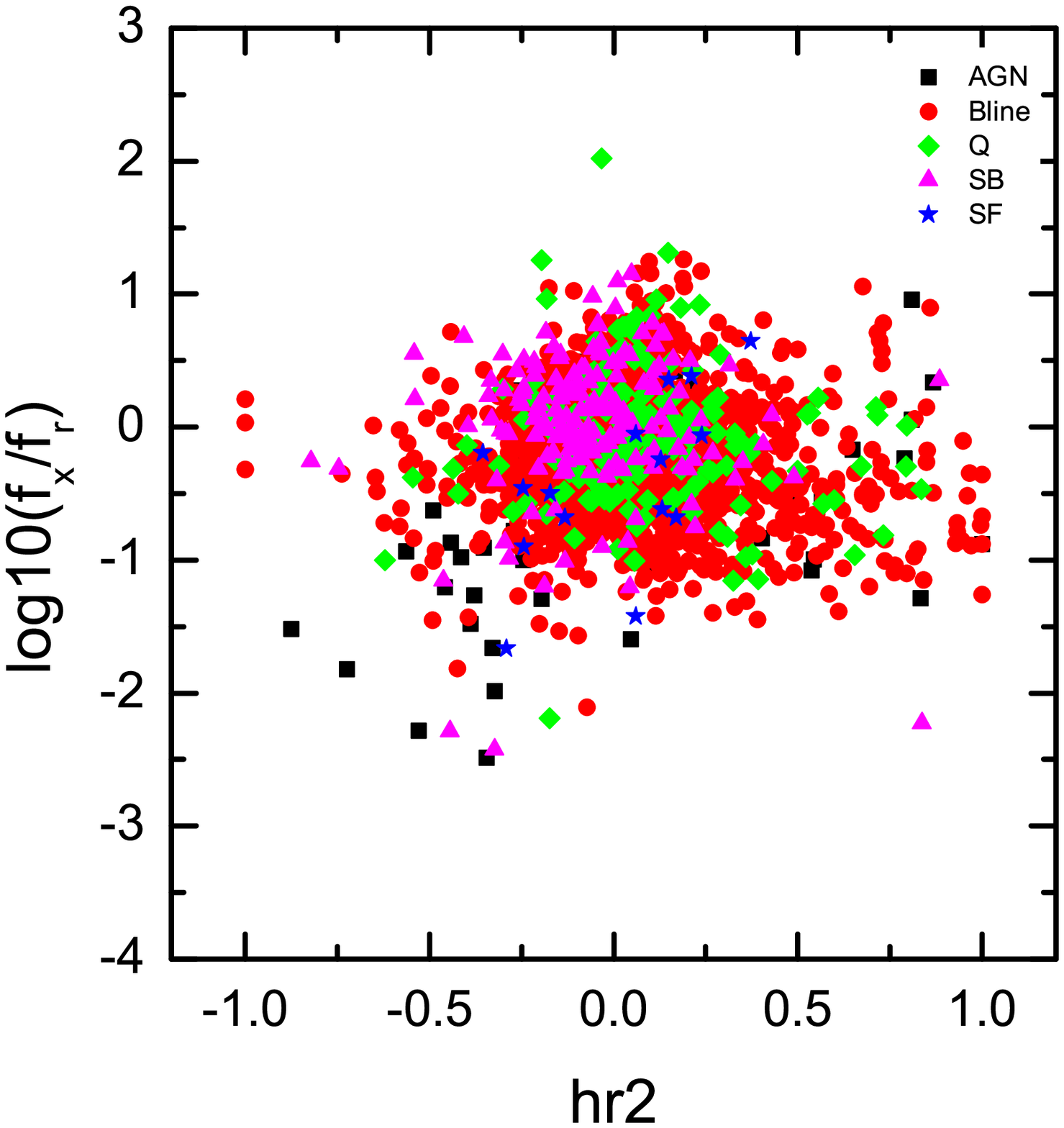}
\includegraphics[bb=11 16 509 522,width=6cm,clip]{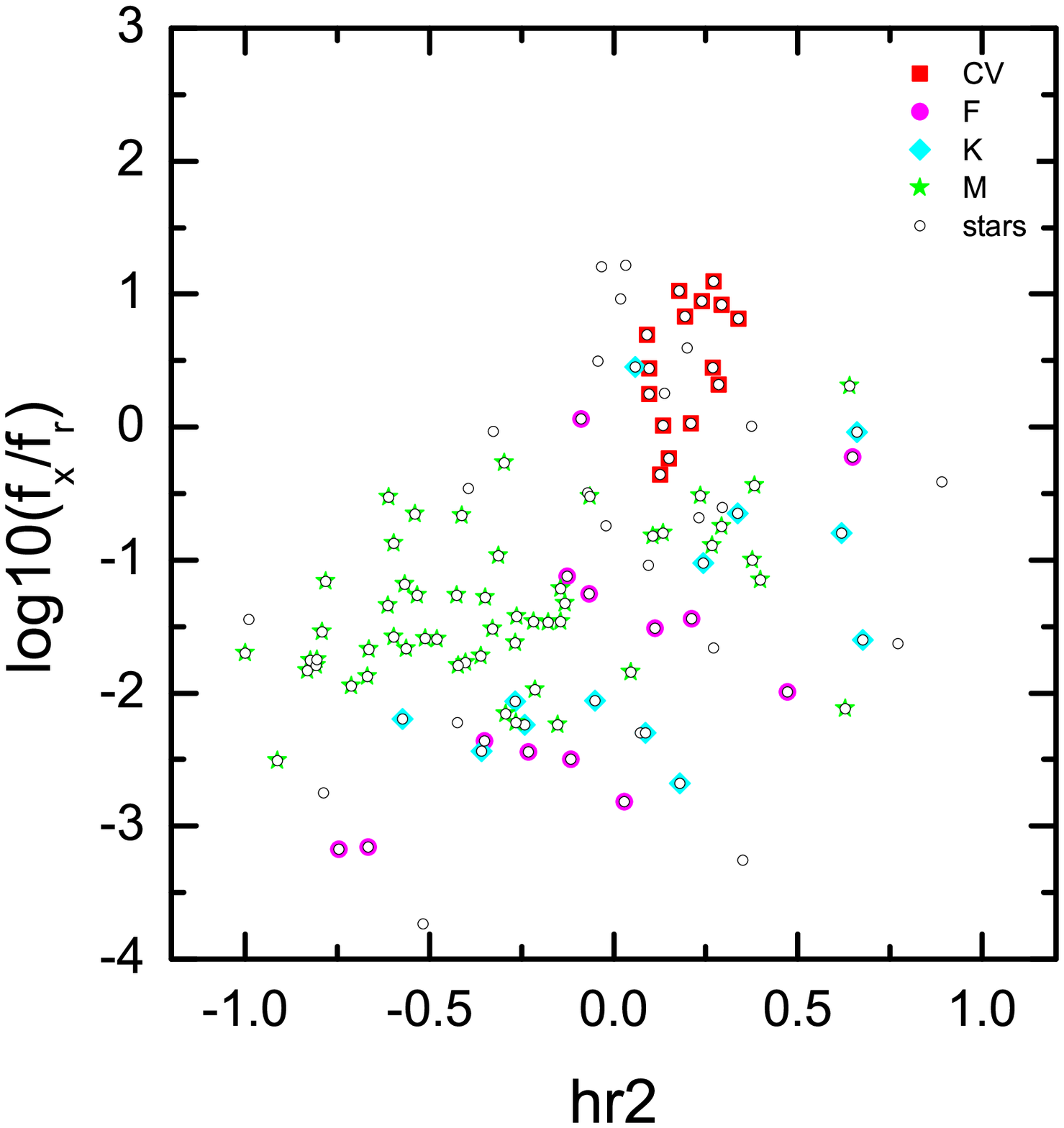}
  \caption{Distribution of sources in the optical band in the log($f_{\rm x}/f_{\rm r}$) versus $hr2$ diagram. The other information is the same as in Fig. 1. }
\end{figure}

\begin{figure}
\includegraphics[bb=11 16 509 522,width=6cm,clip]{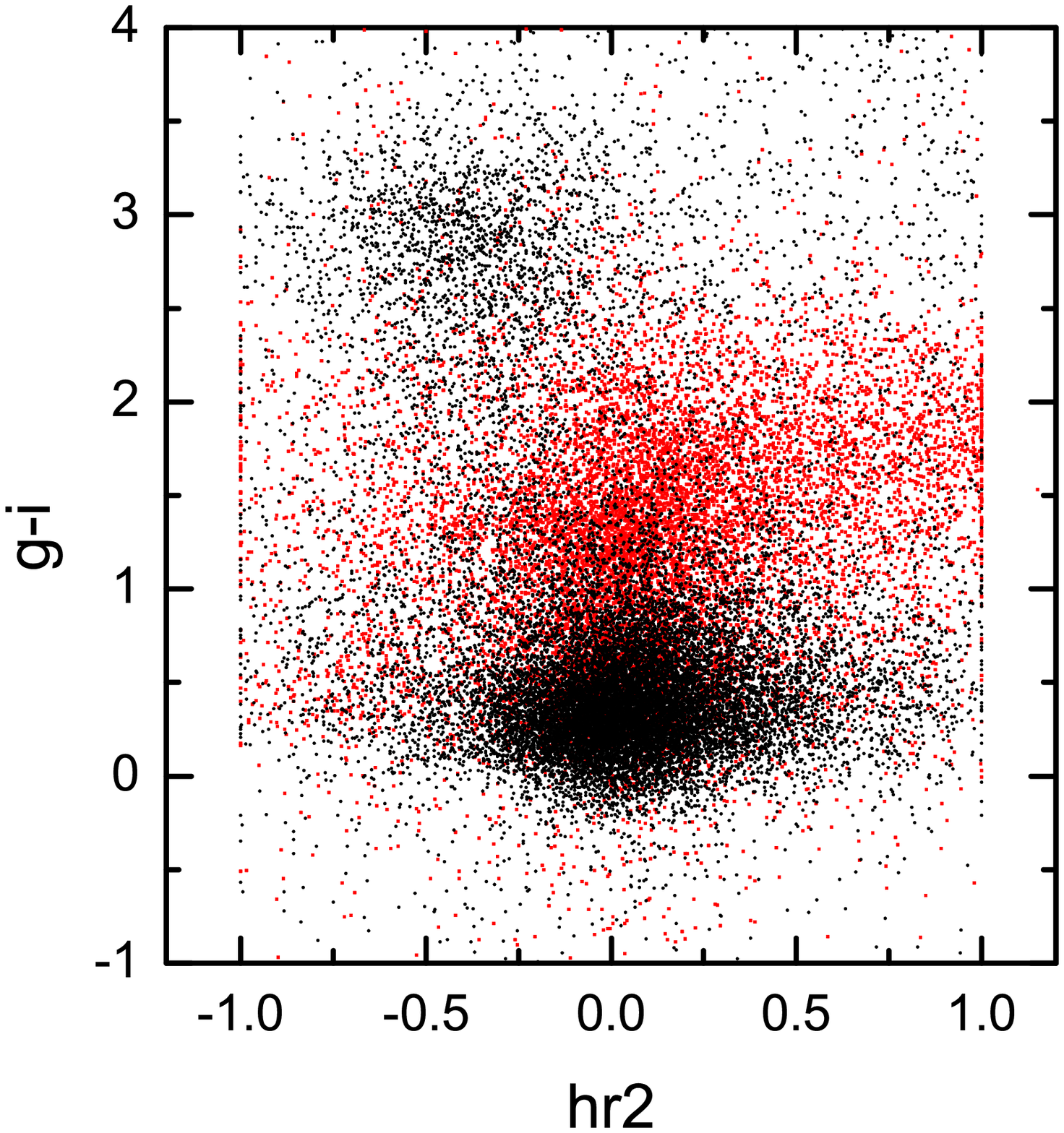}
\includegraphics[bb=11 16 509 522,width=6cm,clip]{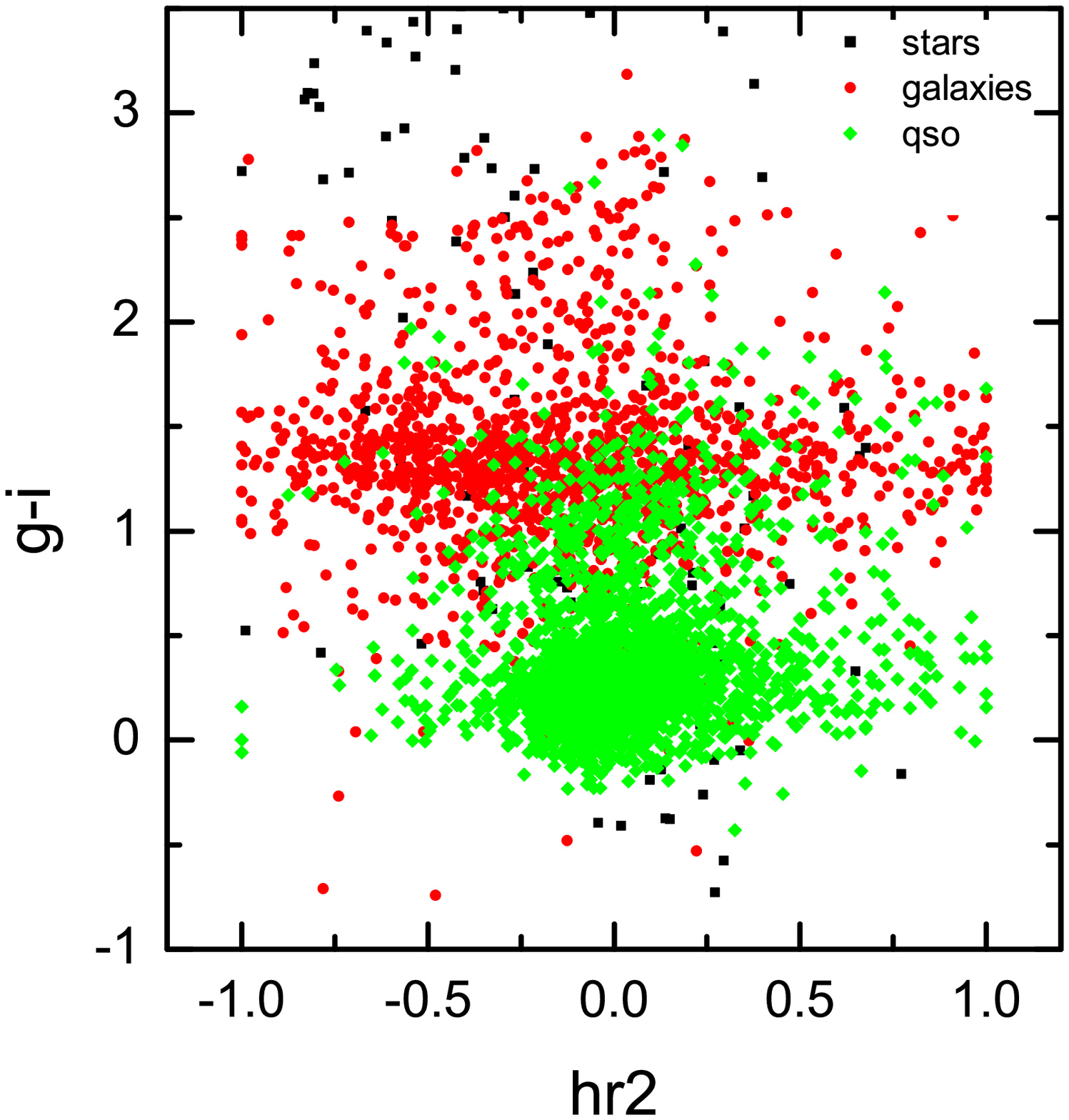}
\includegraphics[bb=11 16 509 522,width=6cm,clip]{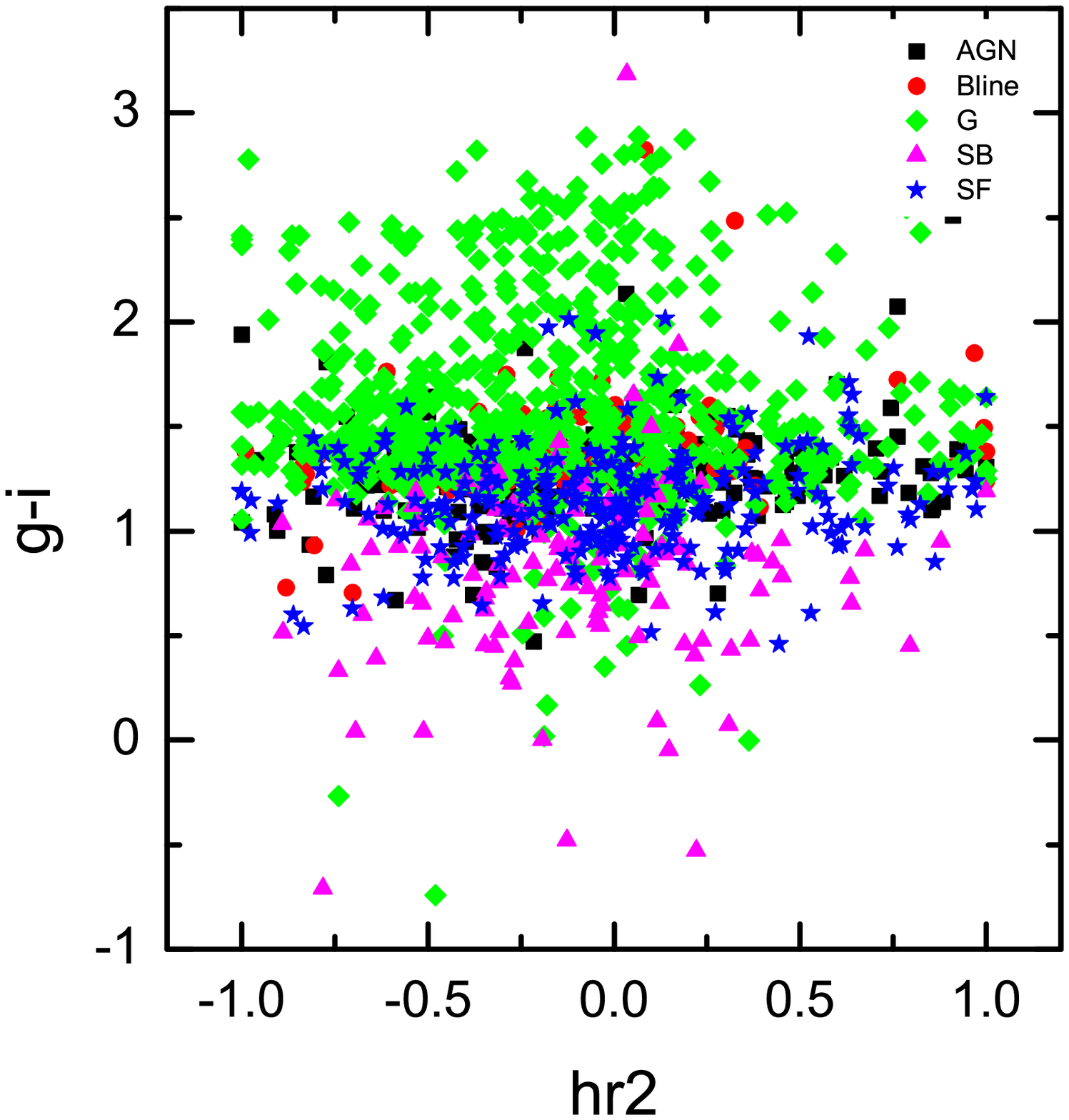}
\includegraphics[bb=11 16 509 522,width=6cm,clip]{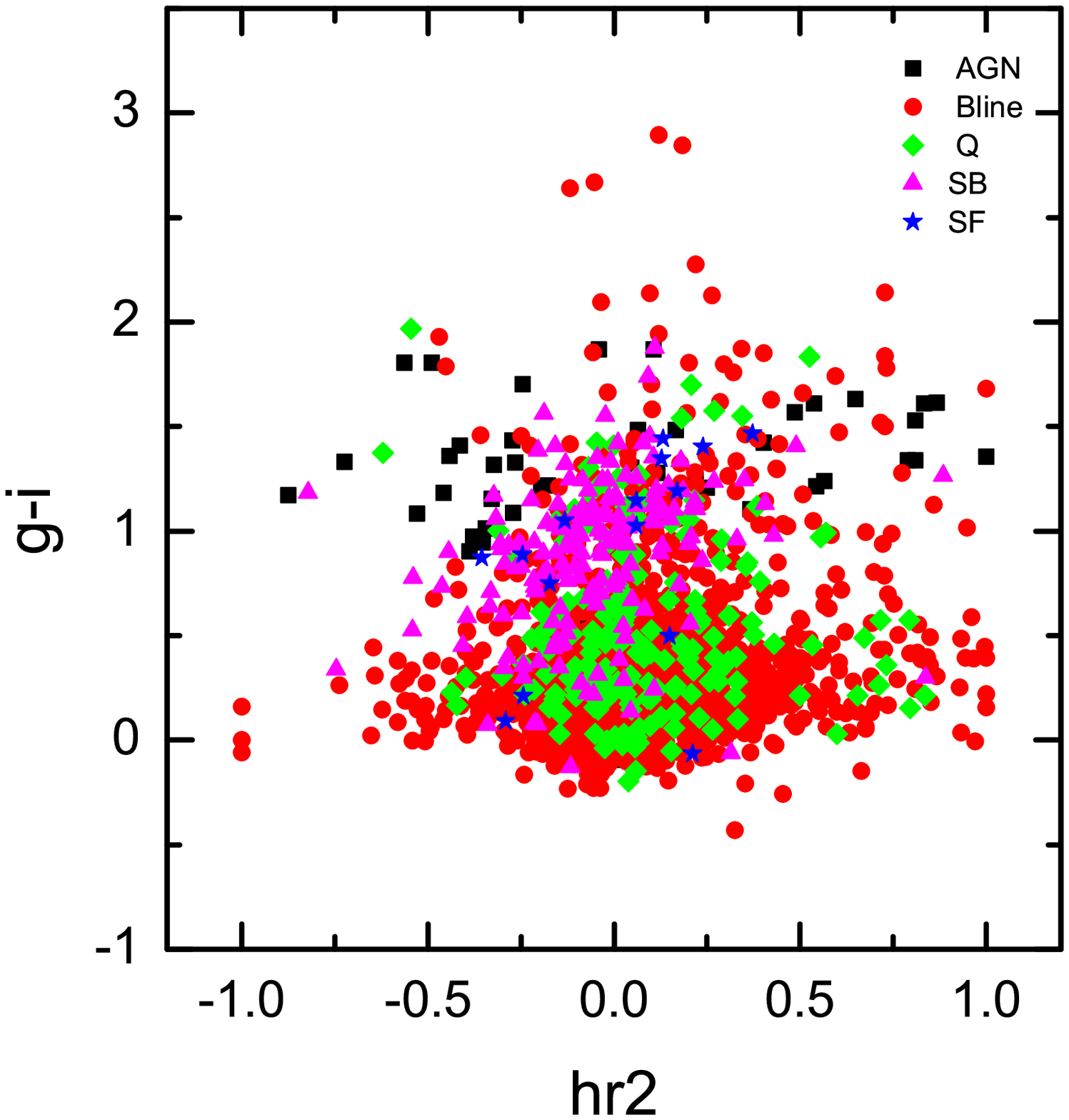}
\includegraphics[bb=11 16 509 522,width=6cm,clip]{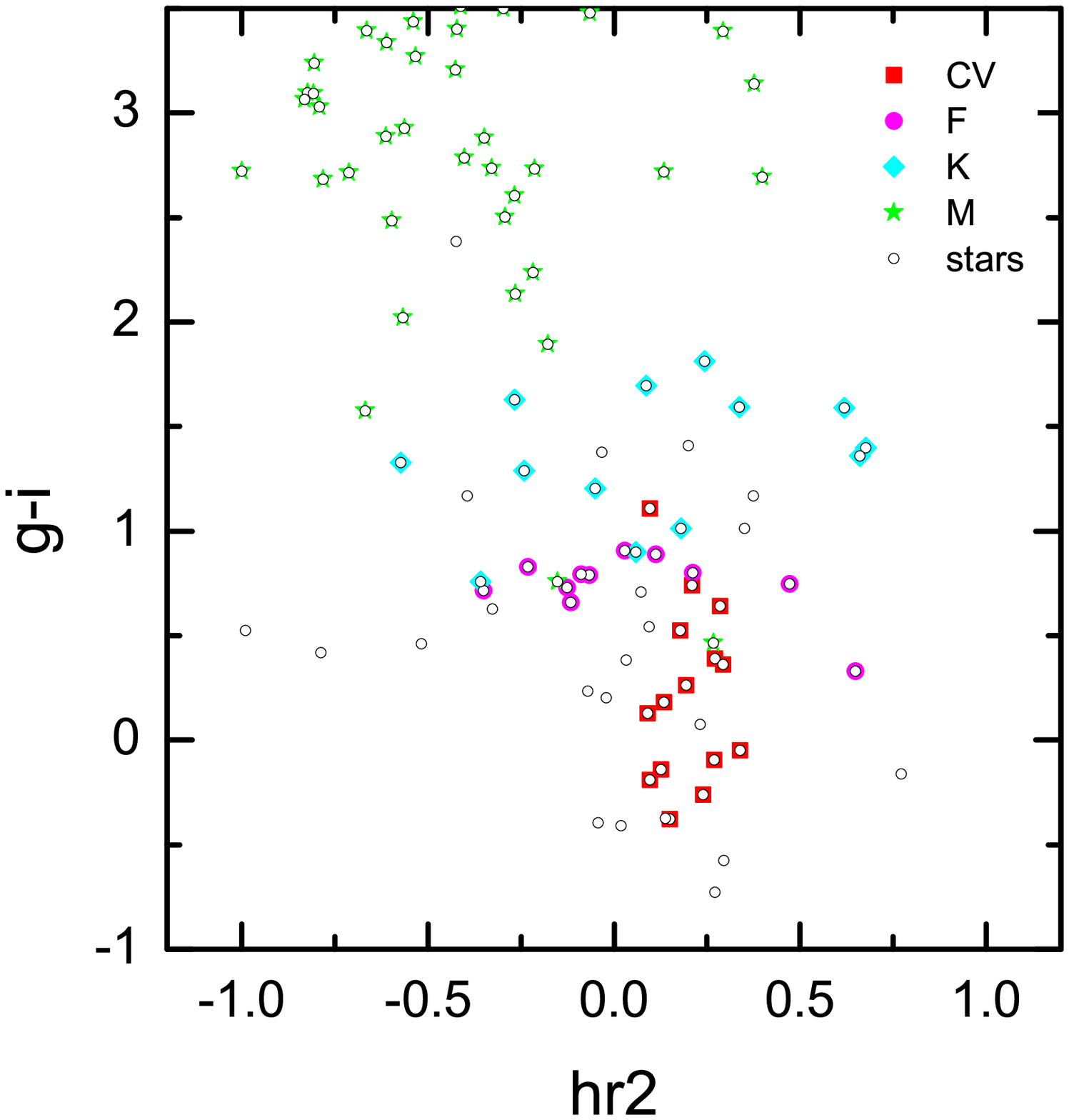}
  \caption{Distribution of sources in the optical band in the $g-i$ versus $hr2$ diagram. The other information is the same as in Fig. 1.}
\end{figure}

\begin{figure}
\includegraphics[bb=11 16 509 522,width=8cm,clip]{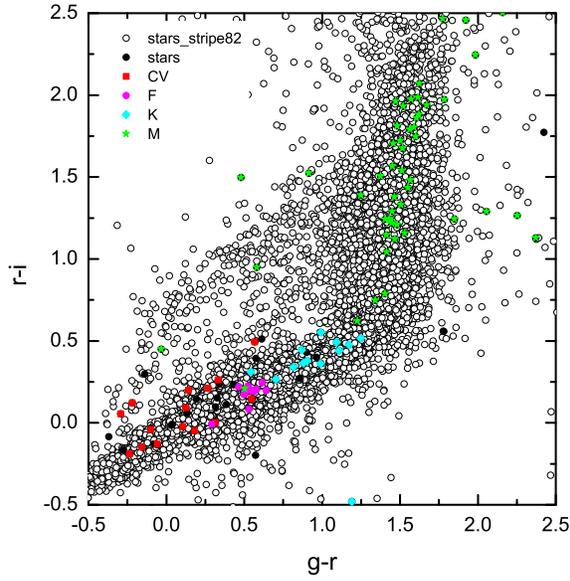}
  \caption{Distribution of stars in $r-i$ versus $g-r$ diagram. Black opened circles: stars from SDSS DR8 stripe 82, filled circles: the whole star sample, red filled squares: CV, magenta filled circles: F stars, cyan filled diamonds: K stars, filled green stars: M stars.}
\end{figure}

\end{document}